# Evolution and breakdown of chiral anomaly in Weyl semiconductor tellurium

Zhifeng Xue[1†], Jiaji Zhao[1†], Yi Ji[1†], Fuyang Chen[2†], Nannan Tang[1], Gangjian Jin[3], Bing Shen[1], Bin Wang[1], Yang Wang[4], Mingxi Chi[4], Jin-Zhu Zhao[2], Ruiqiang Wang[2], Qizhong Zhu[2], He Wang[4], Huakun Zuo[3], Donghui Guo[1], Dingyong Zhong[1], Mingxun Deng[2*], Huichao Wang[1,3*]

[1]Guangdong Provincial Key Laboratory of Magnetoelectric Physics and Devices, Center for Neutron Science and Technology, School of Physics, Sun Yat-sen University, Guangzhou 510275, China

[2]Guangdong Provincial Key Laboratory of Quantum Engineering and Quantum Materials, School of Physics, South China Normal University, Guangzhou 510006, China

[3]Wuhan National High Magnetic Field Center and School of Physics, Huazhong University of Science and Technology, Wuhan 430074, China

[4]Center for Quantum Physics and Intelligent Sciences, Department of Physics, Capital Normal University, Beijing 100048, China

†These authors contributed equally to this work

*Corresponding author.

Email: wanghch26@mail.sysu.edu.cn; dengmingxun@scnu.edu.cn

**Abstract**

Chiral anomaly in topological materials manifests as negative longitudinal magnetoresistance, yet the tunability and evolution of this effect remain elusive. Here we show that doped tellurium serves as a versatile platform to realize a giant negative longitudinal magnetoresistance that is highly tunable via the Fermi level and magnetic field. By moving the Fermi level towards the Weyl point, we observe a pronounced enhancement of negative magnetoresistance, which arises from chiral anomaly and robustly persists up to ambient temperature. Beyond the quantum limit, the system exhibits positive linear magnetoconductance, serving as a compelling transport signature of the chiral anomaly in the lowest Landau level. Notably, in the extreme quantum limit at cryogenic temperatures, the magnetoresistance undergoes a dramatic reversal from negative to positive. This reversal points to a potential field-induced phase transition into a collective electronic state driven by enhanced electron–electron interactions. Our findings provide valuable insights into the unconventional chiral transport and emergent correlated quantum states under extreme magnetic fields.

## Introduction

Magnetoresistance (MR) behavior has long been a central theme and frontier in condensed matter physics as it serves as a simple yet powerful tool for examining the electronic transport properties of solids and uncovering underlying physical mechanisms[1]. Unlike ordinary positive MR caused by the Lorentz-force-induced localization of charge carriers via cyclotron motion, negative MR exhibits an unusual resistance decrease with increased magnetic field, holding great potential for low-dissipation sensors and data storage[2]. However, achieving large tunable negative MR in a pristine solid without intricate interface engineering remains a challenge. In magnetic materials, negative MR usually arises from suppressed spin scattering under a magnetic field[3], involving a complex interplay of charge, spin, and lattice. The microscopic mechanisms are not yet fully understood, hindering precise control. In non-magnetic materials, negative MR is relatively rare. Effects due to the Kondo effect[4] or Anderson localization[5-7] are usually weak, often around 1% or less. In topological insulators, Berry curvature can induce negative MR, but this effect is also weak[8, 9]. In the Weyl system, under parallel electric and magnetic fields, chiral fermion transport between Weyl nodes of opposite chiralities generates an additional electrical conductivity and gives rise to negative longitudinal MR (NLMR) that can be much larger[10-18]. This well-defined mechanism, namely the chiral anomaly, possesses notable advantages due to its sensitivity to the Fermi level position, providing an effective approach for achieving tunable large negative MR effects.

While the chiral anomaly provides a promising route for NLMR, its full manifestation and tunability remain underexplored. In a relatively weak magnetic field regime, the chiral-anomaly-induced NLMR behaves quadratically with the applied magnetic field as demonstrated by theoretical and experimental results[10-17, 19]. In the quantum limit (QL) regime, when charge carriers are confined to the lowest Landau level under sufficiently high fields, a linear NLMR is theoretically proposed as a hallmark of the chiral anomaly[10, 20-22], but this behavior remains to be experimentally validated. In strong magnetic fields, the one-dimensional nature of the carrier energies favors the emergence of electronic instabilities. Especially, the reduction of dimensionality and the high degeneracy in the QL regime can lead to an enhanced electron-electron interaction (EEI), and thus collective transitions are prone to occur[23-26]. For instance, a correlation-driven charge density wave (CDW) state may form, opening a gap around the Fermi level. To thoroughly understand the evolution of chiral-anomaly-induced NLMR and its interplay with correlated effects (Fig. 1A), it is thus crucial to conduct systematic transport measurements at high magnetic fields.

The elemental Weyl semiconductor tellurium (Te) emerges as an ideal platform for these investigations[27, 28]. Its simple, chiral crystal structure hosts well-isolated Weyl points[29, 30], and its semiconducting nature allows for fine-tuning of the Fermi level[31-33]. This tunability, inaccessible in most metallic Weyl systems[34, 35], is central for modulating the contribution of chiral Weyl fermions to transport. Moreover, the QL can be readily achieved in Te at accessible magnetic fields[27], paving the way for probing correlation physics. Here, we report a tunable and giant NLMR in tellurium by modulating the Fermi level and magnetic field synergistically. Specifically, NLMR is more pronounced when the Fermi level of the doped semiconductor is closer to the Weyl node, reaching a maximum -60% and persisting to ambient temperature. The reduction in NLMR as the Fermi level moves further away indicates that the underlying mechanism is related to Weyl physics. Moreover, the NLMR shows a crossover from quadratic to linear dependence when the system enters the QL regime, which well demonstrates the theoretical prediction of chiral anomaly[20]. In the extreme QL, a breakdown of the NLMR occurs, switching to a sharply increased positive MR at low temperatures. This phenomenon points to a potential gap opening associated with a field-induced state driven by EEI within the lowest Landau level. The destruction of NLMR offers valuable insights into the magnetic-field-induced exotic quantum states in the QL regime of Weyl systems.

## Results and Discussion

### Electronic band structure and Weyl physics in Te

Tellurium crystallizes in a chiral structure which lacks inversion symmetry. The trigonal crystal structure and the corresponding Brillouin zone are shown in Fig. 1B. Using first-principles calculations, we illustrate the band structure of Te with spin-orbit coupling in Fig. 1C. In the enlarged view (Fig. 1D), two spin-splitting valence bands cross at two points along the H-L high-symmetry line, resulting in the formation of two Weyl points. One is located at approximately -0.2 eV below the valence band maximum, while the other resides at -0.36 eV. We prepared three batches (B1, B2, B3) of samples with different Fermi levels by adjusting growth parameters. The crystals show good thickness uniformity and feature a long narrow geometry (Fig. S1). Hall measurements reveal *p*-type conductivity in all samples (Fig. S2), revealing linear Hall traces in B1 and B2. By using large-field Hall slope (Fig. S2), the extracted hole densities ($n$) at 2 K are ~$2.3\times10^{18}$ $cm^{-3}$ and ~$3.1\times10^{17}$ $cm^{-3}$, respectively, as shown in Fig. 1E. In the B3 sample, the Hall resistance is nonlinear at low temperatures, driven by the influence of an acceptor impurity band created by Te vacancies, which is a well-documented phenomenon in low-doped Te crystals[32]. By incorporating a two-carrier model fitting, the intrinsic valence-band hole density is extracted to be ~$4\times10^{16}$ $cm^{-3}$. The corresponding Fermi levels estimated from the carrier density via the relation $E_F = \frac{\hbar^2(3\pi^2 n)^{2/3}}{2m^*}$ with an effective mass $m^*$ ~ 0.18 $m_0$[36] are approximately -35 meV, -9 meV, and -2.4 meV, respectively.

The relative positions of the Fermi levels across the different batches are further demonstrated by the quantum oscillations in the transverse MR. Shubnikov–de Haas (SdH) oscillations are observed in samples from batches B1 and B2 (Fig. S3), and the extracted nontrivial Berry phase[33] provides support for the existence of massless Weyl fermions. Under a perpendicular magnetic field, the extracted SdH frequencies are approximately ~55 T and ~26 T for the B1 and B2 batches, respectively. A higher SdH frequency $F$ corresponds to a larger cross-section $S_F$[37], indicating a greater hole density $n$ and thus a lower Fermi level $E_F$ located deeper within the valence band. Based on the Onsager–Lifshitz relation, the carrier density can be estimated via $n = \frac{1}{3\pi^2}\left(F\frac{2e}{\hbar}\right)^{\frac{3}{2}}$ assuming a spherical Fermi surface[38]. The calculated carrier densities for B1 and B2 are $2.3\times10^{18}$ and $7.8\times10^{17}$ $cm^{-3}$, respectively, which are comparable to those obtained from our Hall measurements. For the low-carrier-density B3 samples, SdH oscillations become undetectable due to the heavily depleted 3D Fermi surface. Instead, owing to this remarkably small Fermi pocket, the system enters the extreme QL regime at a low magnetic field threshold below 5 T[36], paving the way for the emergence of log-periodic oscillations (Figs. 1F-1H)[27]. This remarkable phenomenon is associated with the discrete scale-invariant quasi-bound states formed by the relativistic Weyl fermions in a supercritical Coulomb attraction potential[39-42]. Together, these distinct quantum oscillation profiles signify the influence of Weyl physics on the electron transport properties across the three different samples batches.

**Fermi-level-modulation of NLMR**

The longitudinal MR curves measured under the configuration B // I (along *c*-axis) are presented in Figs. 2A-C, for representative samples from each batch. Clearly, SdH oscillations are superimposed on the LMR in both B1 and B2 samples at low temperatures (Fig. 2D). Fast Fourier transform (FFT) analysis reveals two frequencies for each sample when B // I // *c*-axis, consistent with the dumbbell-shaped hole Fermi surface of tellurium that hosts two extremal cross-sectional areas[33]. The medium-carrier-density (B2) sample exhibits frequencies of approximately 8.3 T and 12.5 T, while the high-carrier-density (B1) sample shows frequencies around 21.5 T and 27.6 T. A pronounced NLMR is observed across all sample batches. Crucially, this NLMR is independent of the spatial electrode configuration (Fig. S4) and sample geometry (Fig. S5), effectively ruling out extrinsic artifacts such as current jetting or other accidental effects[43]. In our samples, the onset of the QL regime is achieved when the magnetic field reaches the first frequency, indicating that carriers from this specific extremal orbit have fully entered the lowest Landau level. Notably, the NLMR emerges prominently well below these QL thresholds, excluding the quantum-limit Coulomb-type impurity scattering[44].

Due to the existence of massless Weyl fermions in Te, the observed NLMR likely stems from the chiral anomaly[27]. Experimental studies of the Adler–Bell–Jackiw (ABJ) chiral anomaly in condensed matter can be traced back to vortex experiments in superfluid $^3$He-A in 1997[45, 46], which provided quantitative evidence for anomaly-induced momentum transfer. In subsequent decades, transport signatures of the chiral anomaly have been extensively investigated in solid-state Weyl and Dirac materials[11-13, 15-17, 19, 47-49]. In Te, this assignment is supported by its sensitivity to the relative orientation between the magnetic and electric fields (Fig. S6), a crucial signature aligning with the chiral anomaly[10-18, 50]. At relatively weak magnetic fields, additional magnetoconductivity (MC) driven by the chiral anomaly is expressed as $\sigma_{CA} = C_w B^2$. To achieve a physically complete description of the low-field transport, quantum corrections such as weak antilocalization/weak localization (WAL/WL) and ordinary background MC are incorporated via the expression[14]

$$\sigma_{xx}(B) = C_w B^2 - C_{WAL}\left(\sqrt{B}\,\frac{B^2}{B^2+B_c^2} + \gamma B^2 \frac{B_c^2}{B^2+B_c^2}\right) + \sigma_0. \quad (1)$$

Here $C_w$ and $C_{WAL}$ denote the chiral anomaly and 3D WAL/WL coefficients, respectively. The critical field $B_c$ marks the crossover threshold where WAL/WL correction transitions from $B^2$ to sublinear behavior[51]. We fit the experimental NLMR data prior to the QL using this model, as depicted by the blue dashed lines in Figs. 2A-C. The extracted positive $C_{WAL}$ values (Tables S1 to S3) consistently indicate a robust WAL effect at low temperatures and weak fields, a common observation in Weyl systems due to the nonzero Berry phase of the Weyl nodes[14, 32]. More importantly, the obtained parameter $C_w$ as a function of temperature is plotted in Fig. 2E, in which B3 samples are excluded because their low QL field (~2.5 T at 2K) compresses the valid pre-QL fitting window to a narrow temperature range (≤ 5 K). For the B1 and B2 samples, the temperature evolution of $C_w$ is well-described by the relation[15, 18] $C_W = \frac{e^2}{\pi\hbar}\frac{3}{8}\frac{e^2}{\hbar c}\frac{v_F^3}{\pi^3}\frac{\tau_v}{T^2+\frac{\mu^2}{\pi^2}}$, where $c$, $e$, $\hbar$, $v_F$, $\tau_v$, and $\mu$ represent the speed of light, elementary charge, reduced Planck's constant, Fermi velocity, chirality-changing scattering time, and chemical potential, respectively. Therefore, the observed NLMR profiles are accounted for by the chiral anomaly as the dominant underlying mechanism[27].

To further investigate the NLMR behavior, we analyze additional samples and summarize the coefficient $C_w$ at 2 K in Fig. 2F. A clear modulation of the chiral anomaly is demonstrated by shifting the Fermi level. Crucially, while the raw curves of the B2 and B3 samples display a visually larger low-field NLMR than B1, relying on the visual profile of raw MR curves across different batches is misleading. Because the LMR scales as MR = $1/(\rho(0)C_wB^2+1)-1$, the small zero-field resistivity $\rho(0)$ of the highly conductive B1 sample severely compresses its apparent MR amplitude. The extracted coefficient $C_w$ unambiguously yield the monotonic scaling trend $C_w$(B1) > $C_w$(B2) > $C_w$(B3), which remains robust and independent of the inclusion or exclusion of the WAL/WL terms (Fig. S7). Given the hole-dominated characteristics, the highest carrier density in B1 samples corresponds to a Fermi level closest to the Weyl nodes. The systemic enhancement of $C_w$ as $E_F$ approaches the Weyl points establishes a definitive link between the NLMR and the microscopic Weyl physics, underscoring the Fermi-level-dependent feature of the chiral anomaly in Weyl systems[14, 52].

**Evolution and destruction of NLMR in the QL regime**

To elucidate the evolution of the NLMR in the QL regime, we performed measurements at ultrahigh magnetic fields (Fig. 3A). The NLMR persists from relatively weak fields to the QL regime, showing an abrupt decrease upon reaching the onset of the QL (~21.5 T for batch B1). We fit the low-field data using Eq. (1), and the predicted curve extrapolated to high fields (red dashed lines in Fig. 3B) shows a significant deviation from the experimental results. Instead, the positive MC beyond the QL seems to follow a linear-like dependence. We fit the MC beyond the QL using $\sigma_{xx}(B)=aB^2+bB+c$ as displayed in Fig. 3C. The large ratio $|b|/|a|$ (Fig. 3D) demonstrates that a linear dependence dominates in the QL regime, in contrast to the quadratic dependence at weak fields. This linear trend is inconsistent with an impurity-scattering mechanism, which predicts a quadratic field dependence in the QL limit[44, 53]. Furthermore, the impurity model predicts that the linear coefficient in the QL should be inversely proportional to the carrier density[54]. In our samples, the carrier densities satisfy $n_{B1} > n_{B2}$. Polynomial fittings of the MC for a representative B2 sample in the QL regime are also conducted and shown in Fig. 3E. The larger extracted coefficient for the higher-carrier-density sample (Fig. 3F) contradicts the expectation of the impurity-scattering model, which suggests a smaller coefficient for higher concentrations[44, 54].

The NLMR superimposed with quantum oscillations resembles the behavior of a Weyl phase exhibiting chiral anomaly[20, 55]. According to the theoretical results, the chiral-anomaly-induced positive longitudinal MC (PLMC) can be incorporated into a unified formula[20] $\Delta\sigma = \frac{2e^2}{h}\frac{e^2B^2\tau_v}{\hbar^3}\frac{1}{\varrho(E_F)}$, where the density of states at the Fermi level under Landau quantization is expressed as $\varrho(E_F) = \frac{eB}{h}\frac{1}{h v_F}\left(2\sum_{N=0}^{N_c}\frac{1}{\sqrt{1-2N(\hbar\omega_c/E_F)^2}} - 1\right)$. Here, $N_c = int[\,E_F^2/2(\hbar\omega_c)^2]$, and $\omega_c = v_F\sqrt{|eB|/\hbar}$ is the cyclotron frequency. In the weak-field regime where $E_F \gg \hbar\omega_c$, $\varrho(E_F) \propto E_F$ and $\Delta\sigma \propto B^2$. In the QL limit where $E_F \ll \hbar\omega_c$, $\varrho(E_F) \propto B$ and $\Delta\sigma \propto B$. Consequently, the chiral-anomaly-induced NLMR exhibits a characteristic crossover from quadratic to linear dependence on the magnetic field[20-22]. This prediction remains a challenge to be verified, as it requires a material to reach QL at experimentally attainable magnetic fields being masked by competing mechanisms such as multi-band contributions. In our study, the valence band structure of tellurium near the Fermi surface involves a simple pocket, and the low carrier density enables the full manifestation of the QL transport behavior under strong magnetic fields. Our experimental observation of the linear NLMR beyond the QL thus reveals the dominant contribution of chiral anomaly in the lowest Landau levels.

In the high-carrier-density sample B1-1, the maximum NLMR reaches approximately -60% at high fields, and this behavior robustly persists even at ambient temperature. In the Weyl system, the linearly dispersive band feature of the Weyl node can lead to relatively large Landau level gaps at high magnetic fields[36]. These large gaps effectively suppress thermal excitation and preserves Landau quantization at elevated temperatures, enabling the observation of chiral-anomaly-induced NLMR even at ambient temperatures under large fields.

A striking feature of the Weyl system in the extreme QL regime is the crossover from NLMR to a sharply enhanced positive LMR (PLMR), as shown in Fig. 3A. This feature cannot be attributed to the SdH effect, as detailed analyses of the SdH oscillations and Hall carrier density consistently demonstrate that all carriers are driven into the lowest Landau level beyond 30 T (Supplementary Note 1.1), where no subsequent oscillations can physically occur. This PLMR can arise from a residual perpendicular field component ($B_{\perp}$) caused by a slight misalignment between the sample plane and the applied in-plane magnetic field in samples with extremely large transverse MR[11]. However, in our results, the LMR is even larger than the transverse MR (Fig. S8) and angle-dependent scans reveal that increasing the misalignment angle even makes the crossover field higher (Fig. S6), indicating that the misalignment alone cannot account for the significant LMR change. Quantitatively, in sample B2-8, the transverse MR increases by ~4% whereas the LMR rises by ~38% in the field range of 10-14 T. In sample B2-10, the transverse MR increases by ~35% while the LMR increases by ~57% over the same field range of 10-20 T (Fig. S8). Therefore, the misalignment-induced transverse MR is insufficient to explain the pronounced PLMR in the extreme QL regime. Positive MR can also be driven by quantum interference and orbital effects[56-58], but they are generally dominant at low magnetic fields and insufficient to account for the PLMR observed in the extreme quantum limit (Supplementary Note 1.2). Furthermore, the short-range neutral impurity scattering mechanism[53] can be ruled out, which theoretically predicts a monotonic $B^2$ scaling in contrast to the observed exponential-like growth.

In narrow-gap semiconductors, impurity potential can be crucial and magnetic freeze-out driven by electron-impurity interactions can lead to a reduction in carrier density and an increase in MR[33]. However, in our measurements at ultrahigh magnetic fields, we do not detect obvious changes in either the Hall resistivity or transverse MR (Fig. S8). Previous research has shown that no magnetic freeze-out could be observed at 2 K in undoped tellurium[31, 59], probably owing to differences in impurity type. Furthermore, tellurium has a high dielectric constant, resulting in strong overlap between the wave functions of adjacent impurities. This overlap substantially reduces Coulomb scattering and makes magnetic freeze-out unlikely[60]. Given these observations, the physical origin of the switch from NLMR to PLMR in the extreme QL remains further investigation. It is noted that, for samples within the same batch, minor variations in the absolute values of LMR exist, which can be attributed to their microscopic differences among individual crystals. Despite minor sample-to-sample variations, every single device universally reproduces the evolution from NLMR to PLMR, which is a shared transport characteristic across multiple independent samples (Supplementary Note 1.3).

**EEI-induced CDW under strong magnetic fields**

The switch from NLMR to PLMR occurs within the QL regime, suggesting a possible connection to the strong electronic instabilities under high degeneracy of the lowest Landau level. In the extreme QL, electron motion becomes quantized in the plane perpendicular to the magnetic field, while remaining free along the field direction. This effectively quasi-one-dimensional character of the electronic spectrum promotes the formation of collective states, such as a CDW phase[23-25]. Theoretically, such a transition is predicted to be of a second-order transition[61], exhibiting clear temperature dependence and vanishing at elevated temperatures. The close agreement between our observation and this framework suggestss a plausible description for the observed transport anomalies.

We consider EEI in 3D Weyl semimetals with opposite chirality, which can serve as a reference for the conditions in the doped Weyl system tellurium. In the extreme QL, the effective Hamiltonian around the Fermi level can be written as (see Supplementary Note 2)

$$\widehat{H}^0_{q_z^\chi} = \begin{pmatrix} E_F - \hbar v_F q_z^\chi & \Delta \\ \Delta^* & E_F + \hbar v_F q_z^\chi \end{pmatrix}, \tag{2}$$

where the order parameter is defined as $\Delta = -\left(U_{2k_F}/V\right)\sum_k \langle \Psi^\dagger_{k-2k_F}\Psi^\dagger_k \rangle$, and $\Psi^\dagger_{k\pm k_F}$ and $\Psi_{k\pm k_F}$ denote the creation and annihilation operators for electrons near the Fermi wave vector $k_F$. From Equation (2), the eigenvalue $\varepsilon^\chi_{k_z}$ of the CDW state is obtained as

$$\varepsilon^\chi_{k_z} = E_F - \chi\, sgn\left(q_z^\chi\right)\sqrt{\left(\hbar v_F q_z^\chi\right)^2 + |\Delta|^2}, \tag{3}$$

where $\chi$ represents chirality. The parameter $q_z^\chi$ is defined as $q_z^\chi = k_z - \chi k_F$, and the order parameter $\Delta$ associated with the CDW is solved numerically in a self-consistent way (see Supplementary Note 2). Considering $\chi = 1$, we have $\lim_{k_z \to k_F^-} \varepsilon^\chi_{k_z} = E_F + \Delta$ and $\lim_{k_z \to k_F^+} \varepsilon^\chi_{k_z} = E_F - \Delta$. When $\chi = -1$, we can get $\lim_{k_z \to k_F^-} \varepsilon^\chi_{k_z} = E_F - \Delta$ and $\lim_{k_z \to k_F^+} \varepsilon^\chi_{k_z} = E_F + \Delta$. This means that the lowest Landau level opens a gap at the Fermi level (Figs. 4A, B), with the gap size determined by the magnitude of the CDW order parameter. The gap opening in the lowest Landau level leads to a reduction in conductivity, resulting in the suppression of NLMR.

In this scenario, the total conductivity can be expressed as $\sigma_{zz}(B) = \sigma_D + \Delta\sigma$, with $\sigma_D$ and $\Delta\sigma$ being the zero-field Drude conductivity and magnetic-field-induced conductivity, respectively. Using $E_F$ = 24 meV, the calculated MC (Fig. 4C) reveals a pronounced suppression of NLMR. Figure 4D shows the phase diagram based on the values of order parameter Δ, where the black solid line denotes the phase transition boundary. For weak EEI or relatively weak magnetic fields, the Landau levels cannot open a gap, corresponding to the blue region in Fig. 4D. As magnetic field increases, Δ becomes finite and the system transitions to the CDW state (red region in Fig. 4D). The phase transition threshold corresponds closely to the maximum of the NLMR. Importantly, the simulated phase transition boundary shifts systematically to higher magnetic fields as the temperature increases, matching the qualitative trends of our experimental observations (Fig. 4E). The experimental transport data at different temperatures can be captured by this theoretical model (Fig. 4E). Furthermore, the temperature dependence of the resistivity exhibits metallic behavior prior to the transition field while shifting to insulating behavior afterward (Fig. 4F, Fig. S9), consistent with the physical picture of energy gap formation. The theoretically predicted exponential law ($y=ae^{bx}+c$) further well reproduces the sharply enhanced PLMR (Figs. 4G-I). Therefore, the scenario of a field-induced CDW phase in the QL regime offers a consistent framework to understand the dramatic switch from NLMR to a significantly enhanced PLMR under extreme magnetic fields.

## Conclusion

In summary, this work presents systematic investigations of the electron transport properties of tellurium, uncovering phenomena underpinned by Weyl physics and strong electron correlations. The chiral-anomaly-induced NLMR is shown to be effectively modulated by both the Fermi level position and magnetic field strength. Notably, when the Fermi level shifts towards the Weyl point, the NLMR exhibits enhanced amplitude and stability over a broad temperature range. Furthermore, in the QL regime under high magnetic fields, the NLMR exhibits a characteristic crossover from quadratic to linear field dependence, serving as a hallmark of the chiral anomaly. The high tunability of the NLMR sustained up to ambient temperature establishes tellurium as a promising platform for future device applications. Deep inside the QL regime, the unexpected suppression and reversal of the NLMR point to a potential energy gap opening associated with a field-induced CDW state driven by enhanced EEI. These results highlight the intricate interplay between Weyl physics and many-body correlation effects under strong magnetic fields, advancing the fundamental understanding of quantum transport in Weyl systems.

## Methods

### Crystal Synthesis

The Te crystals were synthesized via the chemical vapor transport method. Firstly, the precursor was prepared by sintering a mixture of Te (3.83 g) and Hf (1.07 g) at 1000 ℃. The obtained precursor mixed with $I_2$ (0.51 g) was sealed in a quartz tube. The hot zone and cold zone of the tube furnace were established at 500 ℃ and 400 ℃, respectively. A growth time of approximately 350 h is typical for medium-carrier-density samples. Increasing the holding time produces samples with a higher carrier density, whereas reducing it produces samples with a lower one.

### Transport Measurements

The electrical transport was measured using a Physical Property Measurement System (14 T-Dynacool PPMS, Quantum Design) and a pulsed high magnetic field facility (Wuhan National High Magnetic Field Center). Gold wires were attached to the clean surfaces of the crystals using conducting silver paste. Standard four-probe method was used to measure MR or Hall resistance.

### Theoretical calculations

First-principles electronic structure calculations were performed using the Vienna ab initio Simulation Package (VASP.6.4.2) along a series of high-symmetry points. The pseudopotentials, representing the ion–electron interactions, were described using the projector-augmented-wave (PAW) method. The exchange–correlation interactions were described by the Perdew–Burke–Ernzerhof (PBE) functional during lattice structure optimization, whereas the Heyd–Scuseria–Ernzerhof (HSE06) hybrid functional was employed for the self-consistent electronic structure calculations. The van der Waals weak interaction in the three-dimensional materials was described within density functional theory (DFT) using the DFT-D3 approach. Convergence tolerances for the total energy and forces were set below $10^{-6}$ eV and $-10^{-2}$ eV/Å for the electrons and ions, respectively. The kinetic energy cutoff for the plane-wave basis was set to 250 eV. During the structural optimization, the Brillouin zones were sampled using a 9×9×7 Γ-centered Monkhorst–Pack (MP) grid. Spin–orbit coupling (SOC) was enabled in the calculations.

**Acknowledgements**

This work is financially supported by the National Natural Science Foundation of China (No. 12374052, 92565303, 12274146, 92165204, 12274145, 12474182, 12574193, 21BAA01133), Guangdong Provincial Quantum Science Strategic Initiative (No. GDZX2401009), the Interdisciplinary program of Wuhan National High Magnetic Field Center (No. WHMFC2025017) Huazhong University of Science and Technology, Guangdong Basic and Applied Basic Research Foundation (No. 2023B1515020050, 20231515010672, 2023B151520013), Guangzhou Basic and Applied Basic Research Foundation (No. 2025A04J5405), Research Center for Magnetoelectric Physics of Guangdong Province (No. 2024B0303390001), Guangdong Provincial Key Laboratory of Magnetoelectric Physics and Devices (No. 2022B1212010008), Guangdong Provincial university Science and Technology Program (No. 2023KTSCX029), the Open Project of the Key Laboratory of Optoelectronic Materials and Technologies (No. OEMT-2023-ZTS-01), and the Instrumental Analysis & Research Center, Sun Yat-sen University.

**Data Availability.** The Source Data underlying the figures of this study are available at https://doi.org/10.5281/zenodo.21881442. All raw data that support the findings of this study are available from the corresponding author upon reasonable request.

**Author contributions**

Z.X., J.J.Z., F.C. and Y.J. contributed equally to this work. H.C.W. conceived and supervised the research. B.S. and B.W. grew the samples; Y.J., G.J., N.T., Y.W., M.C., H.W. and H.Z. carried out the experiments; M.D., F.C., J.Z.Z. and R.W. proposed the theoretical model; H.C.W., Z.X., J.J.Z., F.C, Y.J., M.D., Q.Z., D.G. and D.Z. analyzed the data; H.C.W., Z.X., J.J.Z., F.C, Y.J., M.D., Q.Z., D.G. and D.Z. participated in the discussion; H.C.W., Z.X., J.J.Z., F.C, Y.J. and M.D. wrote and revised the manuscript with comments from all authors.

**Competing interests**

The authors declare no competing interests.

**Additional information**

**Supplementary information** accompanies this paper at the website.

## Figure Captions

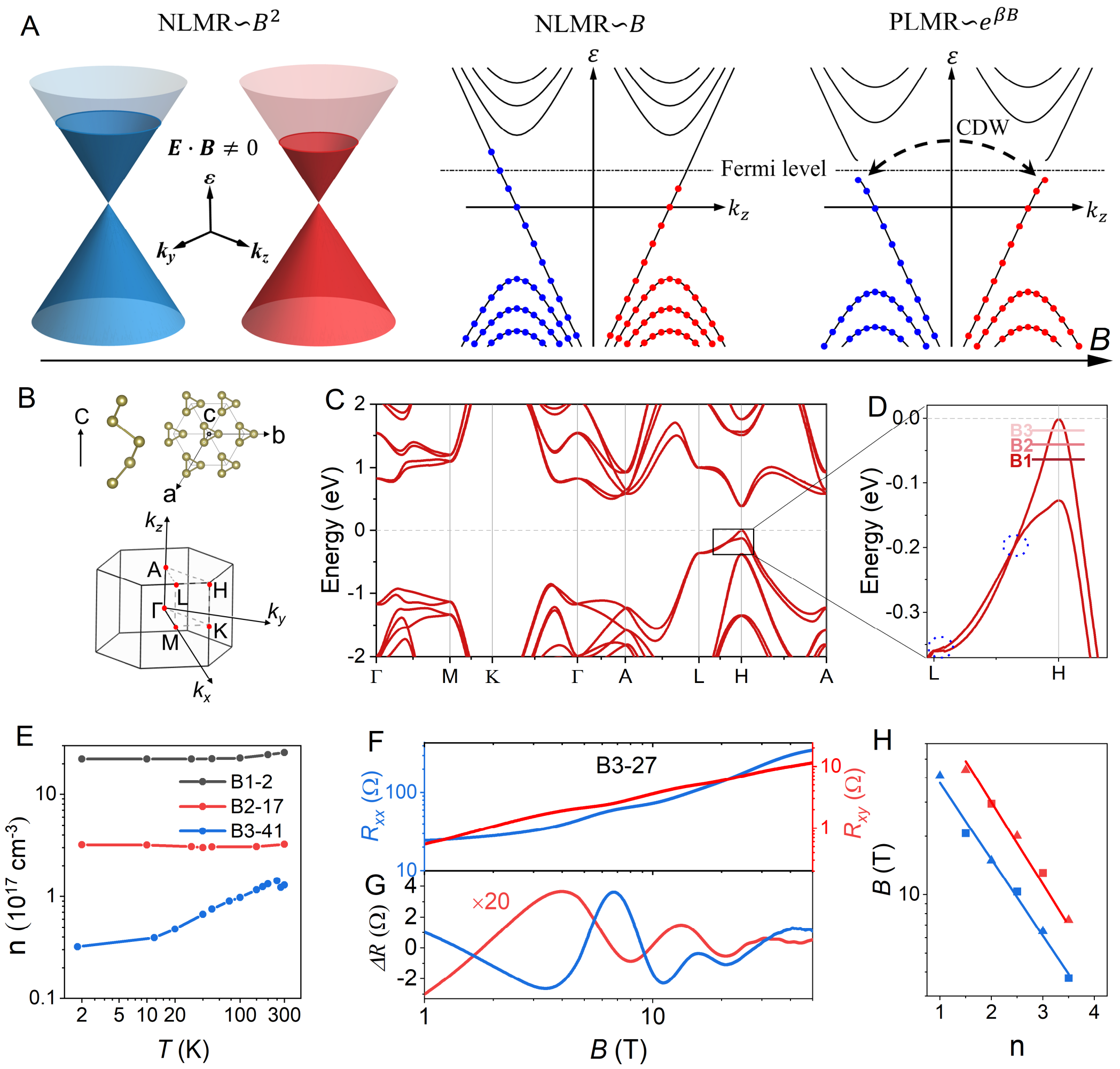

**Figure 1. Electronic band structure and Weyl physics in Weyl semiconductor Te. A** Schematic of the evolution of longitudinal magnetoresistance (LMR) with increasing magnetic field. The negative LMR (NLMR) shows a quadratic dependence on the field at relatively weak fields and transitions to a linear dependence beyond quantum limit when only the lowest chiral Landau level is occupied. A positive LMR (PLMR) emerges in the presence of a charge-density-wave (CDW) gap induced by the correlation effect under strong magnetic fields. **B** Crystal structure and Brillouin zone of trigonal Te. **C** Calculated electronic band structure with spin–orbit coupling. **D** Enlarged view of the boxed region in (**C**) with the Weyl points located along the H-L line. The horizontal lines represent the Fermi levels of samples from three different batches. **E** The typical valence-band hole densities determined from the Hall traces across different batches. Each sample is named by both batch number and sample number. **F** The transverse magnetoresistance and Hall behavior under a perpendicular magnetic field. **G** The oscillation component after subtracting a smooth background. **H** The peaks and valleys are evenly spaced in a logB scale, indicating the appearance of log-periodic quantum oscillations.

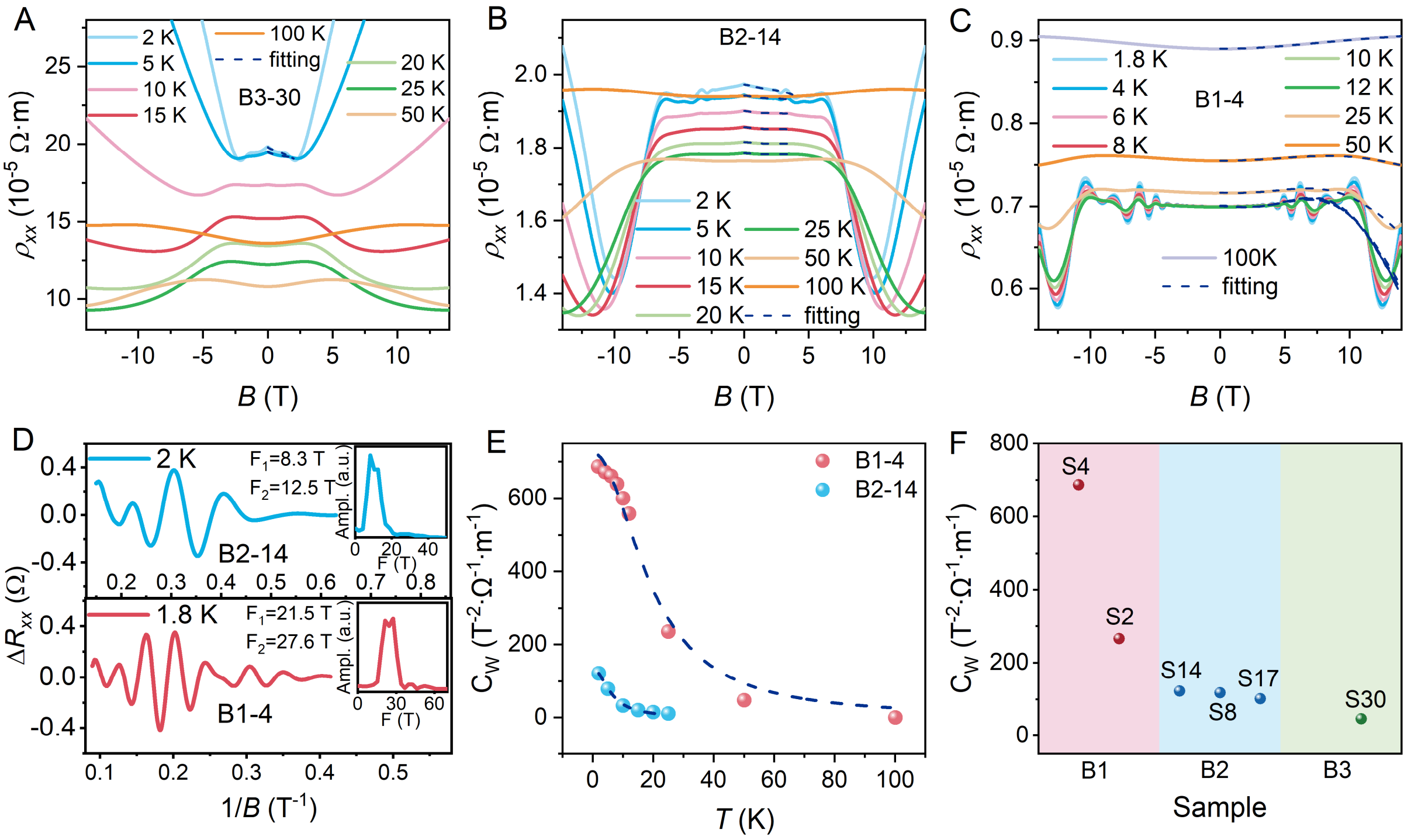

**Figure 2. Fermi-level-modulated negative longitudinal magnetoresistance**. **A-C** The **magnetoresistance** curves ($B$ // $I$ along $c$-axis) of different batch samples. The blue dashed lines represent the fits using $\sigma_{xx}(B) = C_w B^2 - C_{WAL}\left(\sqrt{B}\frac{B^2}{B^2+B_c^2} + \gamma B^2 \frac{B_c^2}{B^2+B_c^2}\right) + \sigma_0$, which incorporates contributions of chiral anomaly, weak antilocalization/localization (WAL/WL) effect and ordinary magnetoconductivity. **D** Shubnikov–de Haas oscillations in the longitudinal magnetoresistance, with the fast Fourier transformation spectrum shown in the inset. The identification of two frequencies is consistent with the dumbbell-shaped hole Fermi surface of tellurium that hosts two extremal cross-sectional areas. **E** Temperature dependence of the chiral anomaly coefficient $C_w$. Dashed lines are the fits by $C_w = \frac{e^2}{\pi\hbar}\frac{3}{8}\frac{e^2}{\hbar c}\frac{v_F^3}{\pi^3}\frac{\tau_v}{T^2+\frac{\mu^2}{\pi^2}}$. **F** Comparisons of $C_w$ for three batches of samples. An enhancement of the coefficient $C_w$ is clearly demonstrated as the Fermi level tunes closer to the Weyl nodes, a trend that remains robust and independent of the exclusion/inclusion of WAL/WL fitting terms.

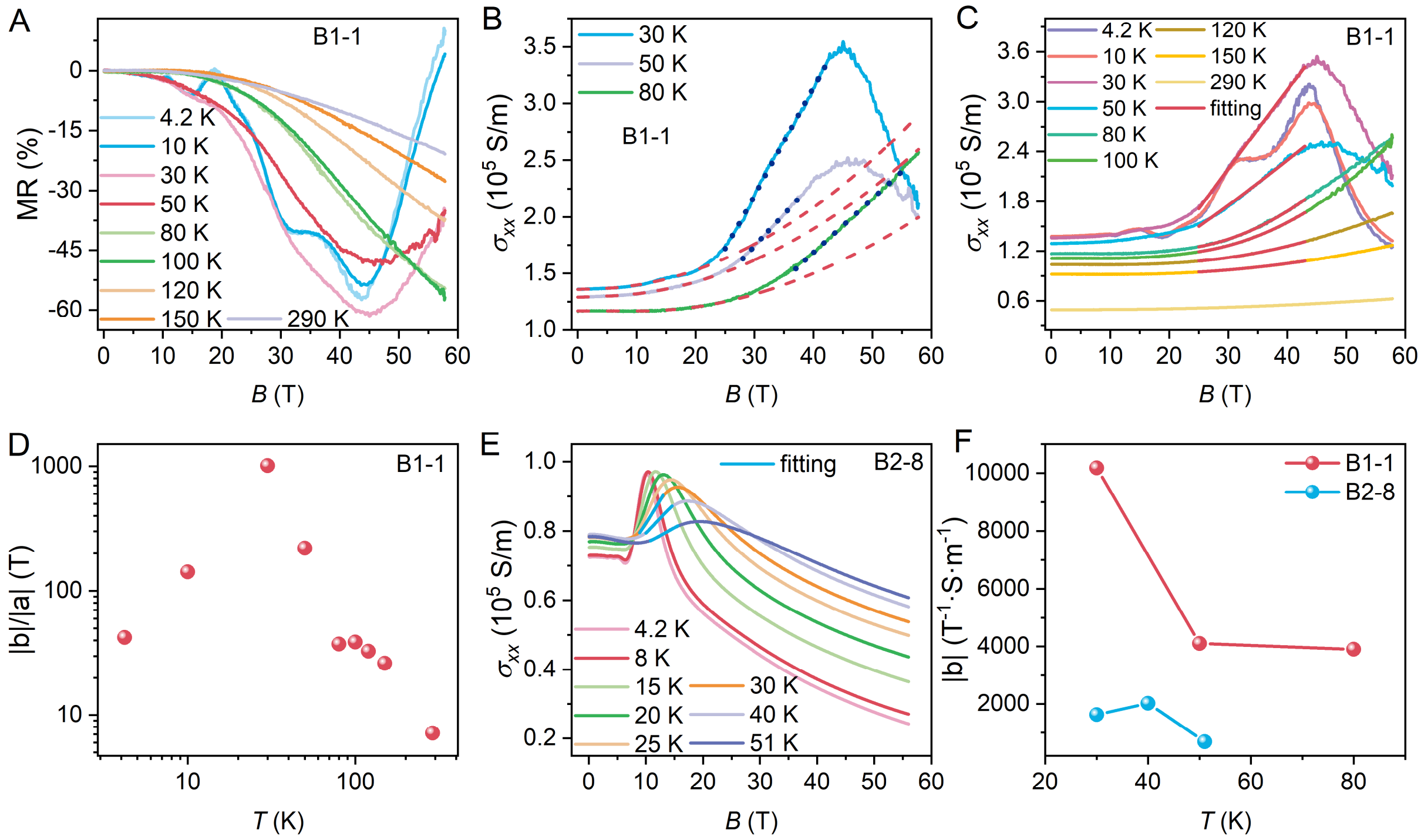

**Figure 3. The evolution of negative longitudinal magnetoresistance under high magnetic fields**. **A** The longitudinal magnetoresistance of sample B1-1 at different temperatures measured under ultrahigh magnetic fields. A switch from negative to positive longitudinal magnetoresistance is observed at low temperatures beyond the quantum limit (QL). **B** The magnetoconductivity (MC) and fits at different temperatures. The red dashed curves represent fits to the low-field data using $\sigma_{xx}(B) = C_w B^2 - C_{WAL}\left(\sqrt{B}\frac{B^2}{B^2+B_c^2} + \gamma B^2 \frac{B_c^2}{B^2+B_c^2}\right) + \sigma_0$. A clear deviation from this fit is observed in the QL regime. At high magnetic fields, the measured MC data exhibits a linear dependence (blue dashed lines). **C** The magnetic field dependence of MC for B1 samples. A polynomial fit was employed on the data in the range of 25-43 T. **D** The obtained parameters in the polynomial fits. It shows that the positive MC is dominated by a linear-$B$ dependence in the QL regime. **E** The MC of sample B2-8 under ultrahigh magnetic fields measured at different temperatures. **F** The fitting parameters from (**C**) and (**E**) show that the higher-carrier-density sample shows larger coefficient b. The linear positive MC in the QL serves as a signature verifying the chiral anomaly inside the lowest Landau level.

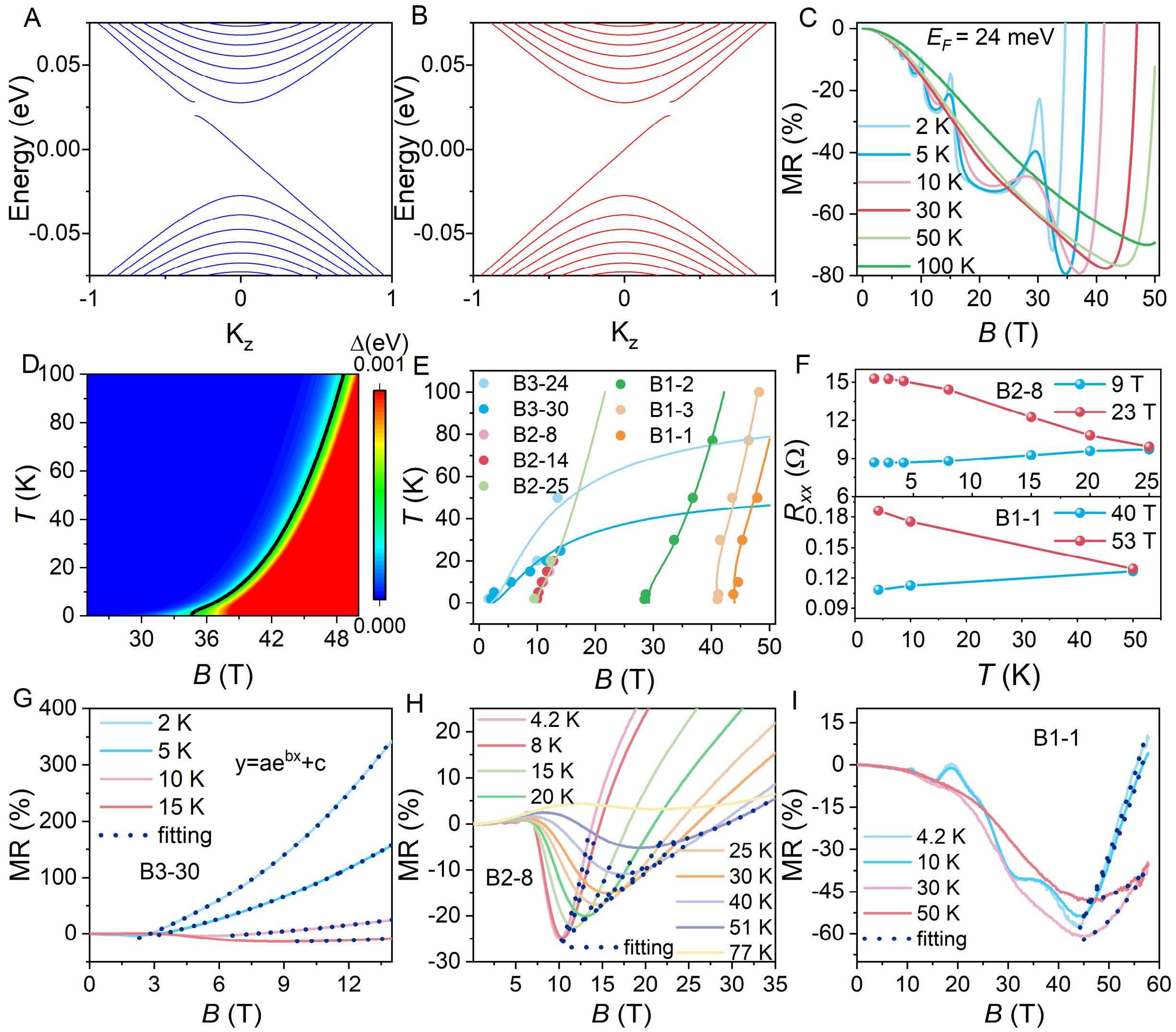

**Figure 4. Suppression and reversal of the negative longitudinal magnetoresistance (LMR) associated with a potential field-induced charge density wave (CDW).** The Landau level in **A** $\chi$=1, and **B** $\chi$=-1 Weyl valleys, in which the lowest Landau level is gapped due to the CDW state near the Fermi level. **C** The emergence of a CDW state induces a switch from negative LMR to positive LMR by numerical simulation at a Fermi level. **D** The calculated phase diagram of the electron-electron interaction-induced CDW phase with varying magnetic fields and temperatures for $E_F$ = 24 meV. The red region shows the band gap opening caused by the formation of a CDW phase. **E** The extracted characteristic fields at which the LMR slope changes sign. The solid lines represent fits using the theoretical model of a transition to CDW state. **F** The temperature-dependent resistance in the positive LMR regime shows an insulating-like characteristic, in contrast to the metallic feature at a relatively weak field. **G-I** The sharply increased positive LMR in the quantum limit regime can be well captured by an exponential fit.